# Dry Glass Reference Perturbation Theory: Development, Applications and Extensions

Bennett D. Marshall*

*ExxonMobil Technology and Engineering Company, 22777 Springwoods Village Parkway, Spring TX 77389 USA*

## Abstract

This manuscript reviews the development, application and extensions of the dry glass reference perturbation theory (DGRPT) closure to the non-equilibrium thermodynamics of glassy polymers (NETGP). DGRPT was developed to allow for the self-consistent and accurate predictions of sorption from complex liquid mixtures into glassy polymers. DGRPT is applied in the context of diffusion theory to predict the membrane based separations of complex liquid mixtures with glassy polymer membranes. Several examples are given, including the membrane based fractionation of crude oil as well as the membrane based separation of highly non-ideal alcohol / hydrocarbon liquid mixtures. Extensions of the theory to higher order expansions are reviewed and evaluated.

*bennett.d.marshall@exxonmobil.com

## I: Introduction

Membrane based separations of liquid mixtures provide a low energy alternative to thermal based separations such as distillation.[1] Organic solvent reverse osmosis (OSRO) and organic solvent nanofiltration (OSN) are membrane based separation of liquid organic mixtures where transport across the membrane is due to molecular diffusion, with no contribution from pressure driven flow through larger pores.[2, 3] Solvents (fluid molecules) are separated based on differences in membrane solubility and diffusivity, Designer polymer membranes allow these solubilities and diffusivities to be tuned based on the molecular structure of the polymer.[4]

High free volume deep glassy polymers are well suited for the separation of liquid mixtures. These polymers provide enhanced stability in organic solvents and allow for faster diffusion due to larger free volume. A challenge with modelling the membranae based separation processes which employ glassy polymers is the solubility predictions of the fluid species within the membrane. The solubility of fluid species in rubbery polymers can be represented by a number of classical equations of state such as Sanchez – Lacomb[5] or PC-SAFT[6]. Glassy polymers are, by definition, not at equilibrium, so these equilibrium methodologies are not directly applicable. The non-equilibrium thermodynamics of glassy polymers (NETGP) of Sarti and Doghieri[7] provides a methodology to extend the equilibrium equations of state to the glassy polymer realm by treating the polymer density as an input to the model, not an output. Simple correlations relating the polymer density to the partial pressure of gas species are enforced on the theory. NETGP has been widely applied to the separation of gases and vapors with glassy polymer membranes. However, imposing a polymer density on the equation of state is more challenging when studying the separation of complex liquid mixtures such as petroleum. For these cases, a minimum of input data is used to parameterize a fluid composed of 1000's of species. The imposed empiricism of the polymer density correlations is not well suited to this task. To allow for the extension of NETGP to these highly complex mixtures, Marshall et al.[8] developed an alternate closure to NETGP. Instead of imposing the polymer density on the theory, the dry glass reference perturbation theory (DGRPT) develops a theoretically self-consistent closure to the theory by expanding the polymer chemical potential around a solvent free reference state. This manuscript reviews the development, application, and extensions of DGRPT.

## II: Membrane Modelling and NETGP

In this section we review the modelling formalism and the assumptions underlying NETGP. Figure 1 gives a diagram of the membrane separation process of a 2 component liquid mixture with a polymer membrane.

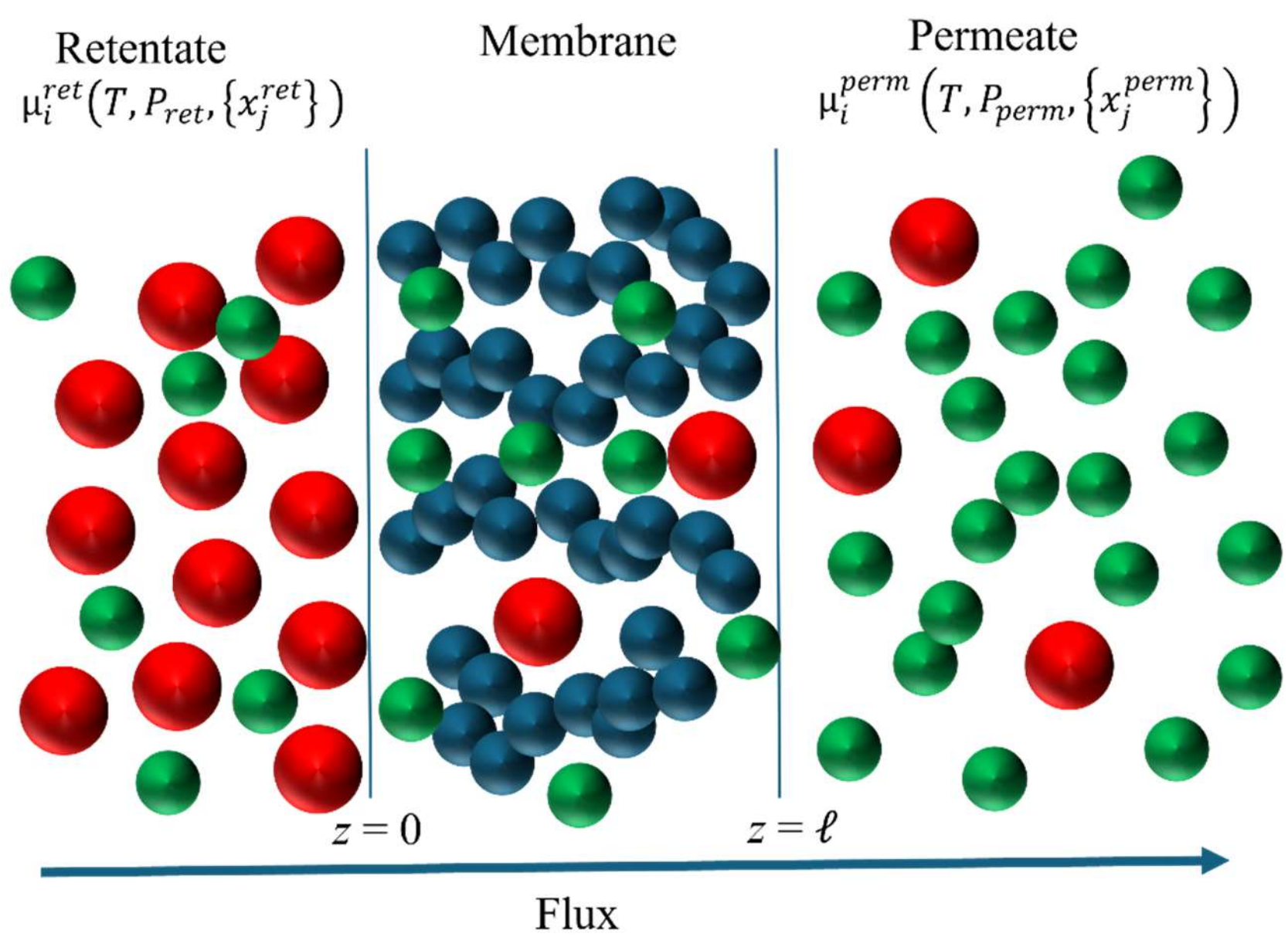

**Figure 1:** Diagram of separation of binary liquid mixture with a polymer membrane

The system is assumed to be isothermal at fixed temperature $T$. The fluid upstream of the membrane is called the retentate and is characterized by the pressure $P_{ret}$, fluid mole fractions $\{x_j^{ret}\}$ and chemical potentials $\mu_i^{ret}(T, P_{ret}, \{x_j^{ret}\})$. The fluid downstream of the membrane is called the permeate and is characterized by the pressure $P_{perm}$, fluid mole fractions $\{x_j^{perm}\}$ and chemical potentials $\mu_i^{perm}(T, P_{perm}, \{x_j^{perm}\})$. The membrane ia a polymer phase which separates the high pressure retentate, from the low pressure permeate. For a molecule to diffuse across the membrane, it must first sorb from the retentate liquid into the polymer at the retentate boundary (z = 0). Once in the membrane, the molecule must diffuse across the membrane to the permeate boundary (z = ℓ). At the permeate boundary, the guest species will be in equilibrium with the permeate.

Multi-component solubility predictions play a central role in the modelling of membrane separation processes. Considering the general case of an $n$ component fluid mixture at a temperature $T$, pressure $P$, set of mole fractions $\{x_j^f\}$. This fluid phase is in equilibrium with a polymer membrane phase which is characterized by $T$, set of sorbed species densities (molecules per volume) $\{\rho_i\}$ and the polymer density $\rho_p$. This equilibrium is expressed by equating chemical potentials in the fluid $\mu_i^f$ and polymer phases $\mu_i^p$.

$$\mu_i^f\left(T, P, \{x_j^f\}\right) = \mu_i^p\left(T, \rho_p, \{\rho_j\}\right) \quad for \quad i = 1{:}n \tag{1}$$

There are $n + 1$ unknown densities in Eqns. (1), but only $n$ equations. Hence an additional "closure" relation is required. At equilibrium this closure relation is provided by enforcing the pressure $P$ in the polymer phase through the following relation

$$P = -\left(\frac{\partial A_p}{\partial V}\right)_{T,\{N_k\}} \tag{2}$$

Where $A_p$ is the Helmholtz free energy of the polymer phase, $V$ is the volume and $\{N_k\}$ are the number of molecules in the polymer phase. As written, the chemical potentials and Helmholtz free energies in Eq. (1) – (2) are described by a density explicit equation of state.

A complication arises when the polymer is in a glassy non-equilibrium state. A glassy state is defined to mean a thermodynamic state which does not correspond to the minimum Gibbs free energy state at constant $T$, $P$. In a glassy state, the pressure relation in Eq. (2) is not valid. Hence, the multi-component solubility problem is not closed. A closure relation to replace Eq. (2) is required.

Sarti and Doghieri[7, 9, 10] proposed the following solution in what is generically called the non-equilibrium thermodynamics of glassy polymers (NETGP). NETGP assumes,

1) The chemical equilibria relations in Eq. (1) hold for the glassy polymer phase.
2) The density explicit equilibrium equation of state can be used calculate the chemical potentials $\mu_i^p(T, \rho_p, \{\rho_j\})$.
3) The closure of Eq. (2) is replaced by imposing the polymer density $\rho_p$ on the theory. The polymer density is no longer an output of the equation of state, but is an input.

The closure relation in NET-GP is typically written as,

$$\frac{\rho_p}{\rho_p^o} = \frac{1}{1+\sum_{i=1}^{n} c_{s,i} f_i} \tag{3}$$

Where $\rho_p^o$ is the "dry glass density", $c_{s,i}$ is an empirical swelling coefficient and $f_i$ is the fugacity of species $i$ in the fluid phase. Equations (1) and (3) summarize the non-equilibrium NETGP flash calculation.

NETGP has been applied to a wide range of systems, but there are features of this theory which may challenge its application to separations of complex liquid mixtures. When dense polymers such as polycarbonate are exposed to gases, they often swell in proportion to the partial pressure (fugacity) of those species. In these dense polymer systems, Eq. (3) provides a physical correct representation of the polymer swelling. However, membranes used in liquid separations typically have a high intrinsic free volume. The high free volume polymers allow for faster diffusion across the membrane. Further, these membranes are typically "deep glasses" with a very high glass transition temperature. The "deep glass" nature allows the membranes to retain free volume when contacted with plasticizing liquids. It has been observed[11] in high free volume glassy polymers that swelling in vapors has a strong non-linear dependence on vapor partial pressure. In fact, high free volume glassy polymers[11] have been shown to be non-swelling at low vapor activities, and linear swelling at high vapor activities. Equation (3) cannot represent this behavior. More recently Merlonghi et al.[12] applied NETGP to predict acetone-methanol mixture sorption in Matrimid membranes. To match the non-linear swelling behavior, a third parameter was introduced to replace linear swelling dependence with power law swelling. The updated density relation is given by

$$\frac{\rho_p}{\rho_p^o} = \frac{1}{1 + \sum_{i=1}^{n} c_{s,i}\, a_i^{n_i}} \tag{4}$$

Where $a_i$ is the activity ($f_i / f_{i,ref}$) and $c_{s,i}$ and $n_i$ are adjustable constants. With Eq. (4), NETGP is a 3 parameter model to describe sorption of a single component (two swelling parameters and 1 binary interaction parameter).

The additional parameters introduced in the empirical density relations Eq. (3) – (4) will challenge the applicability to complex liquid mixtures such as petroleum. Generally, only pure vapor sorption data are available for a few select hydrocarbons from which a full model of petroleum consisting of 1000's of components must be developed. This task is made even more challenging by the introduction of empirical swelling constants for each component as required in Eqns. (3) – (4).

To decrease the number of parameters in NETGP, and better match sorption induced glass transitions, the NETGP-RS was proposed by Minelli and Doghieri[13]. The polymer density in the NETGP-RS is calculated by interpolating between the density calculated from the equilibrium equation of state, $\rho_p^E$, and the dry glass density $\rho_p^o$ through the following relationship

$$\frac{1}{\rho_p} = \frac{1}{\rho_p^o} + \chi\left(\frac{1}{\rho_p^{EQ}} - \frac{1}{\rho_p^{EQ,o}}\right) \tag{5}$$

Where $\rho_p^{EQ}$ is the polymer density obtained by solving the corresponding equilibrium solubility problem with Eqns. (1) – (2) and $\rho_p^{EQ,o}$ is the density of the dry polymer calculated using the equilibrium equation of state. $\chi$ is the ratio of the experimentally measured dry glassy polymer compressibility $\kappa_g^o$ to the equilibrium compressibility calculated with the equation of state $\kappa_{eq}$

$$\chi = \frac{\kappa_g^o}{\kappa_{eq}} \tag{6}$$

In Eq. (6) both compressibilities are evaluated at the glass transition pressure of the pure polymer.

NETGP-RS has been shown to accurately represent pure component gas / vapor sorption in dense glassy polymers such as polycarbonate and polymethylmethacrylate where sorption induces a transition from glassy to equilibrium behavior.[13, 14] However, the application of NETGP-RS to deep glassy polymers used for liquid separations may be challenging. Generally, $\kappa_g^o$ is unknown for most polymers. Further, the method relies on the polymer having an equilibrium equation of state developed from equilibrium data to calculate $\rho_p^{EQ}$ and $\kappa_{eq}$. This is not possible for deep glassy polymers, as rubbery polymer data do not exist. Doghieri attempted to apply NETGP-RS to vapor sorption in the polymer PIM-1[14], but as will be shown, the model gives strange predictions at high vapor activity for several hydrocarbon species.

## III: Dry Glass Reference Perturbation Theory

In an attempt to develop a theoretically self-consistent closure relation to NETGP, Marshall et al.[8] proposed the dry glass reference perturbation theory (DGRPT). Four fundamental assumptions underly the development of DGRPT

1) As in traditional NETGP, Eq. (1) holds true and the fluid species chemical potentials in the glassy polymer phase may be calculated using an equilibrium density explicit equation of state
2) In addition to sorbed fluid species in the polymer, the polymer chemical potential $\mu_p$ may be evaluated with the same equilibrium density explicit equation of state as the fluid species
3) When fluid species sorb into glassy polymers which are deep glasses, these polymers will swell, but the underlying structure of the dry polymer remains imprinted on the swollen polymer
4) As a result of 3), the polymer chemical potential can be expanded around a dry glass reference state to develop a self-consistent closure to NETGP

Assumptions 3) and 4) will be most accurate for deep glassy polymers which retain their glassy structure as they swell, staying far away from glass transitions. On the other hand, assumptions 3) and 4) will be least accurate for dense glassy polymers in the presence of plasticizing gases which induce a transition to a rubbery state. It should be noted that membranes used for liquid separations are typically deep glassy polymers.

When truncated at first order, the DGRPT expansion for sorption from a multi-component fluid is given as

$$\mu_p\left(\rho_p,\{\rho_j\},T\right) \approx \mu_p\left(\rho_p^o,T\right) + \sum_{j=1}^{n} \left.\frac{\partial \mu_p\left(\rho_p^o,\{\rho_k\},T\right)}{\partial \rho_j}\right|_{\{\rho_k\}=0} \rho_j \qquad (7)$$

In Eq. (7), $\mu_p\left(\rho_p,\{\rho_j\},T\right)$ is the polymer chemical potential evaluated at the true polymer density $\rho_p$ and set of sorbed fluid species densities $\{\rho_j\}$, $\mu_p\left(\rho_p^o,T\right)$ is the polymer chemical potential of the dry glass reference state and the derivative in the last term is evaluated in the dry glass limit. The specific form of Eq. (7) results from enforcing the following limit on the polymer density

$$\left.\frac{d\rho_p}{d\rho_i}\right|_{\{\rho_k\}=0} = 0 \qquad (8)$$

Equation (8) enforces the experimental observation that polymer swelling is small at low sorbed fluid densities in high free volume glassy polymers.[11] It should be noted that DGRPT does predict polymer swelling, just not at the infinite dilution limit,

Equation (7) provides a theoretical closure to the NETGP method. There are several potential benefits of this approach. No additional parameters are introduced beyond those used in the equation of state. The extension from pure fluid sorption to multi-component required no additional assumptions. The closure references only polymer phase densities. This is in contrast to Eq. (3), which is directly proportional fluid phase fugacities. As will be shown in section V, relating the polymer density directly to fluid phase fugacities can result in anomalous predictions for systems exhibiting strong competitive sorption.

The dry glass density itself is a function of temperature and pressure, which is described by the following relation.

$$\rho_p^o = \rho_p^{oo}\left(1 + \frac{P}{\varepsilon_y} - \alpha(T - T_o)\right) \qquad (9)$$

where $\rho_p^{oo}$ is the history-dependent dry glass density at atmospheric pressure and $T = T_o$, $\varepsilon_y$ is the Young's modulus of the dry polymer, and $\alpha$ is the thermal expansion coefficient of the dry polymer. Note, unlike Eq. (3) where pressure acts to swell the polymer due to sorbed species, the pressure dependence in Eq. (9) acts to compress the dry polymer. At low pressures, the pressure dependence of Eq. (9) can be neglected, however at the high pressures observed in membrane separations, the pressure dependent term in Eq. (9) can significantly affect sorption predictions.

All chemical potential terms in Eq. (7) are calculated with the same equilibrium equation of state used in Eq. (1). In all previous applications of DGRPT, the PC-SAFT equation of state has been used. The Helmholtz free energy is given by,

$$a = \frac{A}{Nk_BT} = a_{hc} + a_{at} + a_{as} + a_{dp} \tag{10}$$

In Eq. (10) $a$ is the reduced Helmholtz free energy, $N$ is the number of molecules and $k_B$ is Boltzmann's constant. The hard chain[15] free energy contribution $a_{hc}$ is parameterized by the segment number $m$ and segment diameter $\sigma$, and is evaluated with the simplified approach of von Solmes et al.[16] The square well attractive contribution $a_{at}$ includes the square well depth $\varepsilon$, and is evaluated with the PC-SAFT[6, 17] of Gross and Sadowski. The association contribution of the theory $a_{as}$ is evaluated using Chapman's[15] general first order solution (TPT1) to Wertheim's thermodynamic perturbation theory[18-20]. The specific association parameterization Marshall[21] is used. Each molecule is assigned a number of acceptor association sites $n_a$ and number of donor sites $n_d$ based on molecular structure. Each associating species is then parameterized by the association energy $\varepsilon_{AB}$ and association bond volume $\kappa_{AB}$. Finally, the polar contribution $a_{dp}$ is evaluated with the Jog and Chapman[22] polar theory. The polar terms is parameterized scheme of Marshall et al.[23] who described the theory in terms of the combined parameter "polar strength" $\alpha_p$ $= m x_p \mu^2$, where $x_p$ is the fraction of polar segments and $\mu$ is the dipole moment. The cross parameters for two interacting species $i$ and $j$ are given by the standard combining rules[17],

$$\varepsilon_{ij} = \sqrt{\varepsilon_{ii}\varepsilon_{jj}}\left(1 - k_{ij}\right) \tag{11}$$

$$\sigma_{ij} = \frac{\sigma_{ii} + \sigma_{jj}}{2} \tag{12}$$

$$\varepsilon_{AB}^{ij} = \frac{\varepsilon_{AB}^{ii} + \varepsilon_{AB}^{jj}}{2} \tag{13}$$

$$\sigma_{ij}^3 \kappa_{ij} = \sqrt{\sigma_{ii}^3 \kappa_{AB}^{ii} \sigma_{jj}^3 \kappa_{AB}^{jj}} \tag{14}$$

where $k_{ij}$ is the binary interaction parameter. When using DGRPT, the sorption of a fluid phase species in the polymer is parameterized by the binary interaction parameter with the polymer $k_{i,p}$

SAFT equations of state for non-polar molecules are typically parameterized to pure component vapor pressure and liquid density data to determine the pure component parameters $m$, $\sigma$, and $\varepsilon$.[24] Without the inclusion of vapor pressure, or some other measure of liquid cohesive energy, it is not possible to extract an optimal non-degenerate parameter set. As an example, we consider the case of the pure component PC-SAFT parameterization of octacosane. The model is parameterized in two ways which are highlighted in Table 1.

**Table 1:** Two parameterization strategies for octacosane

| Set # | Regress $\rho$ | Regress $P_{sat}$ | $m$ | $\sigma$ | $\varepsilon / k_B$ (K) |
|---|---|---|---|---|---|
| Set 1 | Yes | No | 12.71 | 3.792 | 248.24 |
| Set 2 | Yes | Yes | 11.29 | 3.948 | 253.6 |

In Set 1, only liquid density ($\rho$) data is included in the parameter regression. In Set 2, both liquid density and vapor pressure ($P_{sat}$) are included in the regression. The $\varepsilon$ of each parameter set is similar, but the segment diameter is very different between the two cases. In PC-SAFT, the model predictions are extremely sensitive to the value of the hard sphere diameter. Figure 2 compares liquid density, vapor pressure and heat of vaporization predictions at saturation for octacosane using both parameter sets. Each parameter set accurately represents the density, while only Set 2 accurately represents heat of vaporization and vapor pressure.

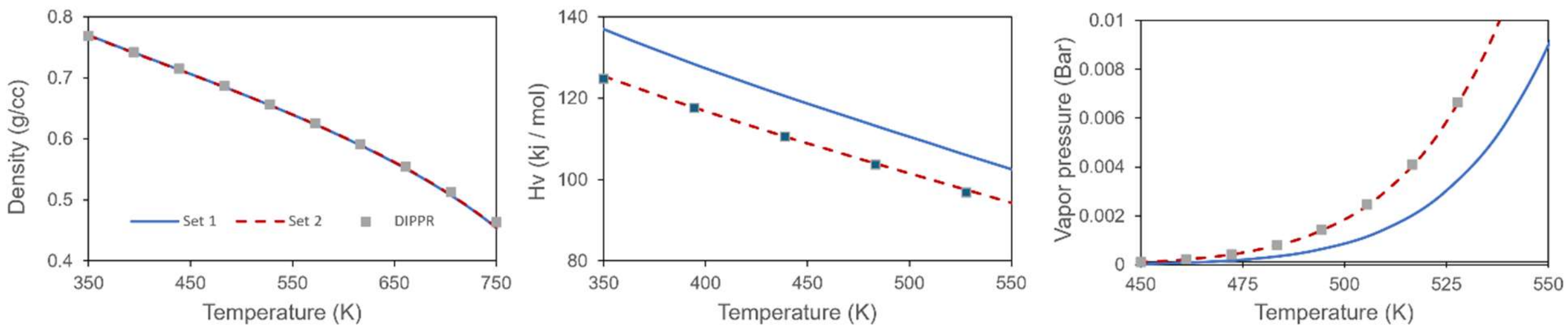

**Figure 2:** Comparison of PC-SAFT predictions of liquid density, heat of vaporization, and vapor pressure at saturation for pure octacosane using parameters Set 1 and Set 2. Symbols represent DIPPR[25] correlation values

Parameter sets which give nearly identical density predictions can give vastly different predictions of other properties which depend on the cohesive energy of the liquid. This exercise demonstrates that liquid density data alone is not sufficient to extract a unique $m$, $\sigma$, $\varepsilon$ which represents both the density and the cohesive energy of the polymer. For this reason, it is common to include binary solubility data in parameter regression. This solubility data takes the place of vapor pressure data which is unavailable for polymers. Very recently, Ismaeel et al. published an in depth review of polymer parameterization methodology.[26]

For deep glassy polymers there is no equilibrium density data to include in the regression of the polymer PC-SAFT parameters. For these cases, our typical approach has been to include vapor sorption data from two fluid species as the basis for the regression of $m$, $\sigma$, and $\varepsilon$ of the polymer. With the polymer pure component parameters fixed, the solubilities of additional fluid species are then parameterized by a single binary interaction parameter $k_{i,p}$ between the solvent species and the polymer. As a reminder, binary interaction parameters are used in the combining rule for the square well attractive energy in Eq. (11). The utility of this approach is then verified, by using these same polymer parameters to predict vapor and liquid sorption from additional species not included in this initial parameterization.

## IV: Application to deep glassy polymers

Assumptions 3) and 4) in the development of DGRPT are most accurate when applied to deep glassy polymers, which have a very high glass transition temperature. Further, the enforcement of Eq. (8) is most accurate when applied to high free volume polymers. For these reasons, the first order DGRPT expansion is most accurate when applied to high free volume deep glassy polymers. This class of polymers is extensively used to develop membranes for liquid phase separations.

### poly[(tri-methylsily) propyne] (PTMSP)

Marshall et al.[8] applied DGRPT to predict multi-component vapor and liquid sorption in PTMSP. PTMSP is a high free volume deep glassy polymer, meeting assumptions 3) – 4) in the development of DGRPT. For PTMSP, the pure polymer parameters $m$, $\sigma$ and $\varepsilon$ were fit to reproduce pure vapor sorption of both dimethyl carbonate (DMC) and methanol. Figure 3 compares DGRPT predictions of the sorption from a binary vapor of methanol and DMC to the data of Vopicka et al.[11, 27]. Activity in the vapor phase is given as the ratio of partial pressure to vapor pressure. Measurements and calculations are performed for several fixed mole fractions of methanol in the vapor phase. DGRPT gives a good representation of this binary system. The model accurately predicts the sorption of each component, and gives a reasonable prediction of polymer swelling as a function of activity and composition. Polymer swelling is represented as the polymer volume $V$, normalized by the dry polymer volume $V_{dry}$.

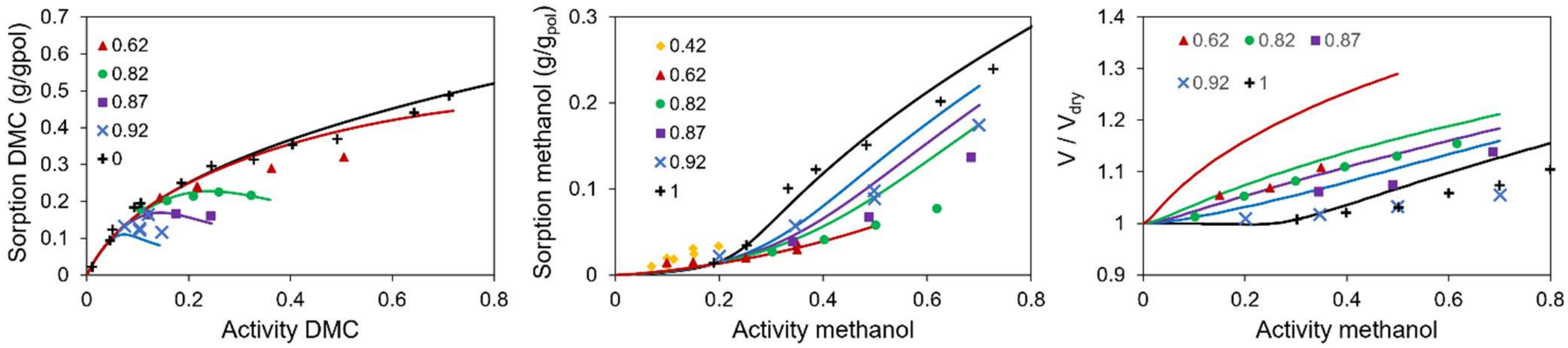

**Figure 3:** Comparison of DGRPT model predictions (curves)[8] to data (symbols) for sorption[27] and swelling[11] of a binary DMC / methanol vapor on PTMSP at 40 °C. Different curves represent different methanol mole fractions in the vapor phase

Further, the DGRPT model[8] was applied to predict polymer swelling in ethanol / water binary liquid mixtures in Fig. 4. Fluid / polymer binary interaction parameters of $k_{i,p}$ = -0.0594 and 0 for ethanol and water respectively were fit to vapor sorption data. The DGRPT calculation for polymer swelling from the binary liquid represents a pure prediction of the model. Note, in agreement with the data, the model predicts that the polymer contracts for small mole fractions of ethanol. This accurate theoretical prediction provides strong empirical justification for DGRPT in the description of sorption in the high free volume glassy polymer PTMSP. The traditional NETGP theory with the empirical closure Eq. (3), would be incapable of making this prediction.

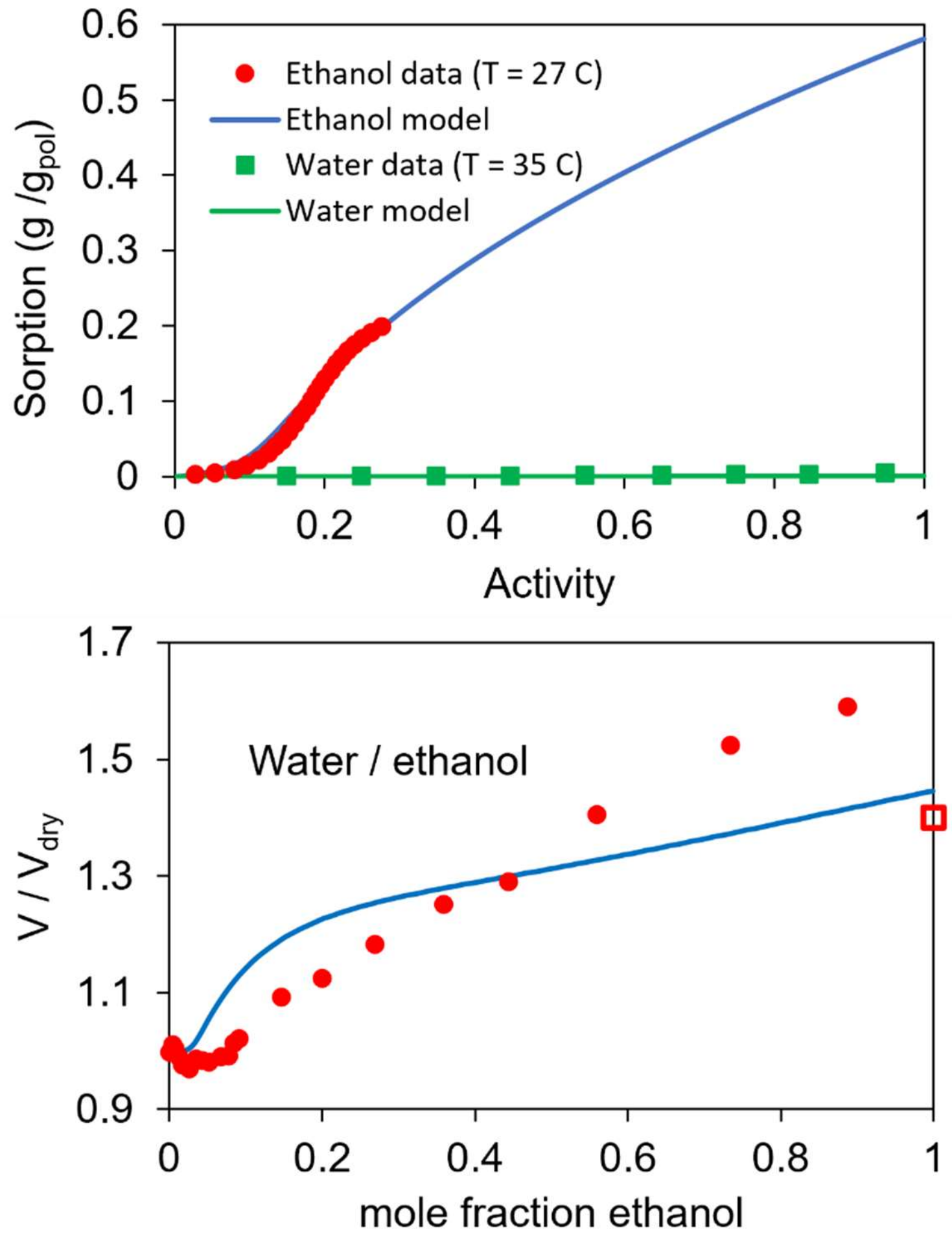

**Figure 4:** Top: Comparison of data (squares-water[28], circles-ethanol[29]) to DGRPT predcitions[8] for pure vapor sorption in PTMSP. Bottom: Comparison of data (circles[28], square[30]) to DGRPT predictions for the swelling of PTMSP in contact with an ethanol / water liquid mixture

**PIM-1**

PIM-1 is a high free volume glassy polymer[31] which has been extensively studied[32] for gas, vapor and liquid separations. DGRPT was applied to predict vapor and liquid solubilities of heptane, toluene, DMC and methanol in PIM-1. In addition, a Maxwell-Steffan diffusion model employing DGRPT solubilities was applied to predict the pervaporation separation of DMC and methanol.

Figure 5 compares DGRPT and NETGP-RS predictions to experimental data for sorption from pure heptane, toluene, methanol and DMC vapors and liquids in PIM-1. As can be seen, DGRPT is highly accurate for the prediction of heptane sorption over the full activity range. When applied to toluene, DGRPT is highly accurate at 55 °C for all activities, but less accurate at high activities at 25 °C. At these high activities the sorption isotherm takes on a linear character. In addition to hydrocarbons, DGRPT is accurate in the representation of sorption from pure methanol and DMC vapors.

The dashed curves represent the NETGP-RS predictions of Doghieri[14]. Both NETGP and DGRPT are single parameter models for these binaries. Each model is parameterized by the binary interaction parameter $k_{i,p}$. NETGP-RS predictions are not accurate for sorption of these vapors in PIM-1. The NETGP-RS transitions to equilibrium type sorption behavior at too low of vapor activities. While NETGP-RS has been shown to work well for gas sorption[14], application of this method to vapor sorption in deep glass polymers is challenging due to the lack of a definite equilibrium state.

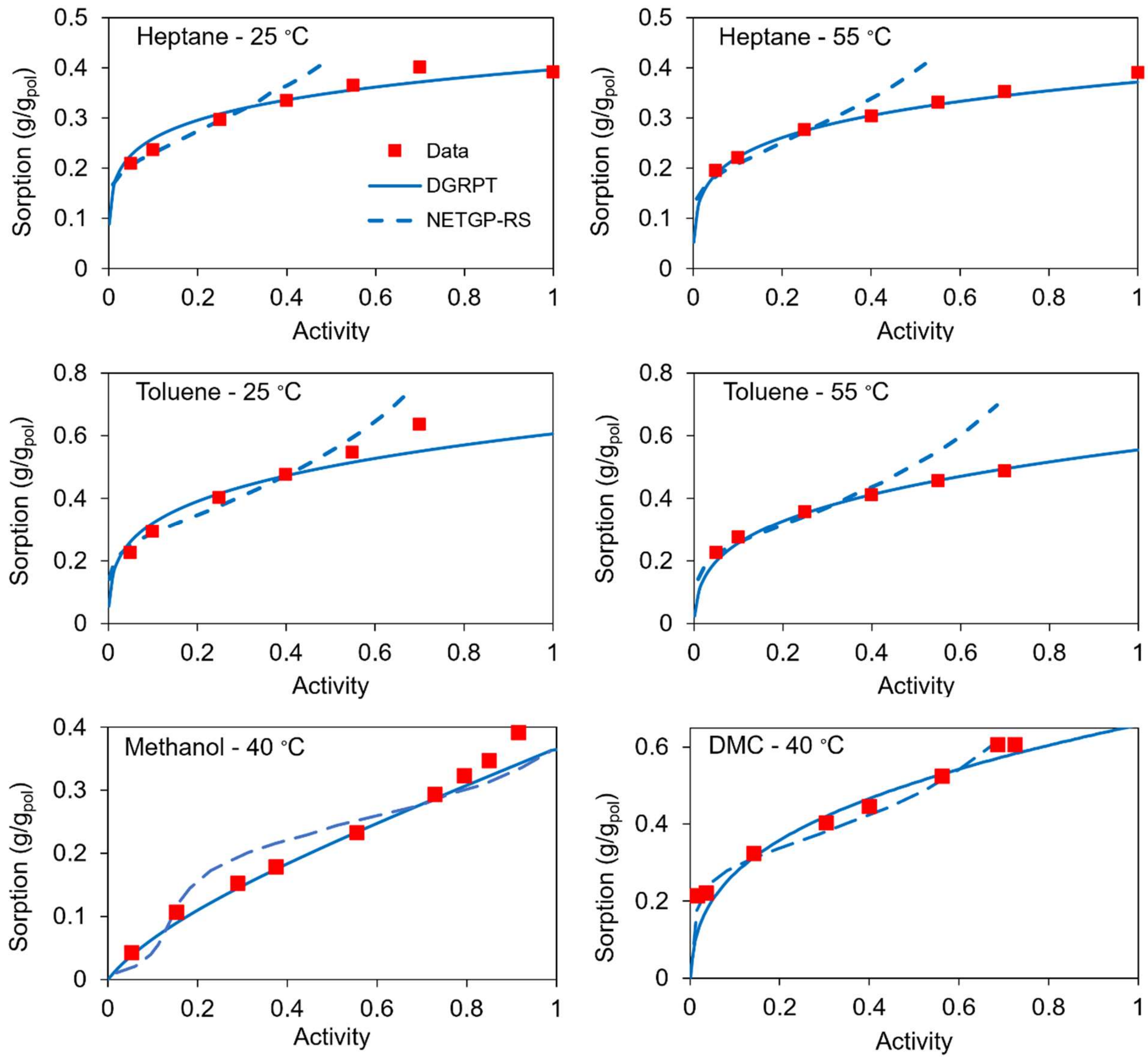

**Figure 5:** Comparison of DGRPT predictions[33] (solid curve), NETGP-RS[34] predictions (dashed curve) to experimental data[35, 36] for the sorption of heptane, toluene, methanol and DMC vapors in PIM-1.

Figure 6 focuses on the application of DGRPT to methanol / DMC mixtures.[33] The top panel compares DGRPT predictions to data for sorption from binary vapor mixtures. As can be seen, DGRPT accurately predicts the mixture sorption. DGRPT + Maxwell-Stefan[37] diffusion was then applied to predict the separation of methanol / DMC liquid mixtures using membrane based pervaporation. The bottom panel of Fig. 6 compares model and data for the separation factor versus methanol concentration in the feed. The separation factor is defined as $\beta_{ij} = x_i^{perm} x_j^{ret} / x_j^{perm} x_i^{ret}$. It was shown that the solubility controls the shape of the separation factor versus composition curve.[33] The DGRPT predictions of mixture solubility are highly accurate, resulting in accurate predictions in the compositional dependence of the separation factor.

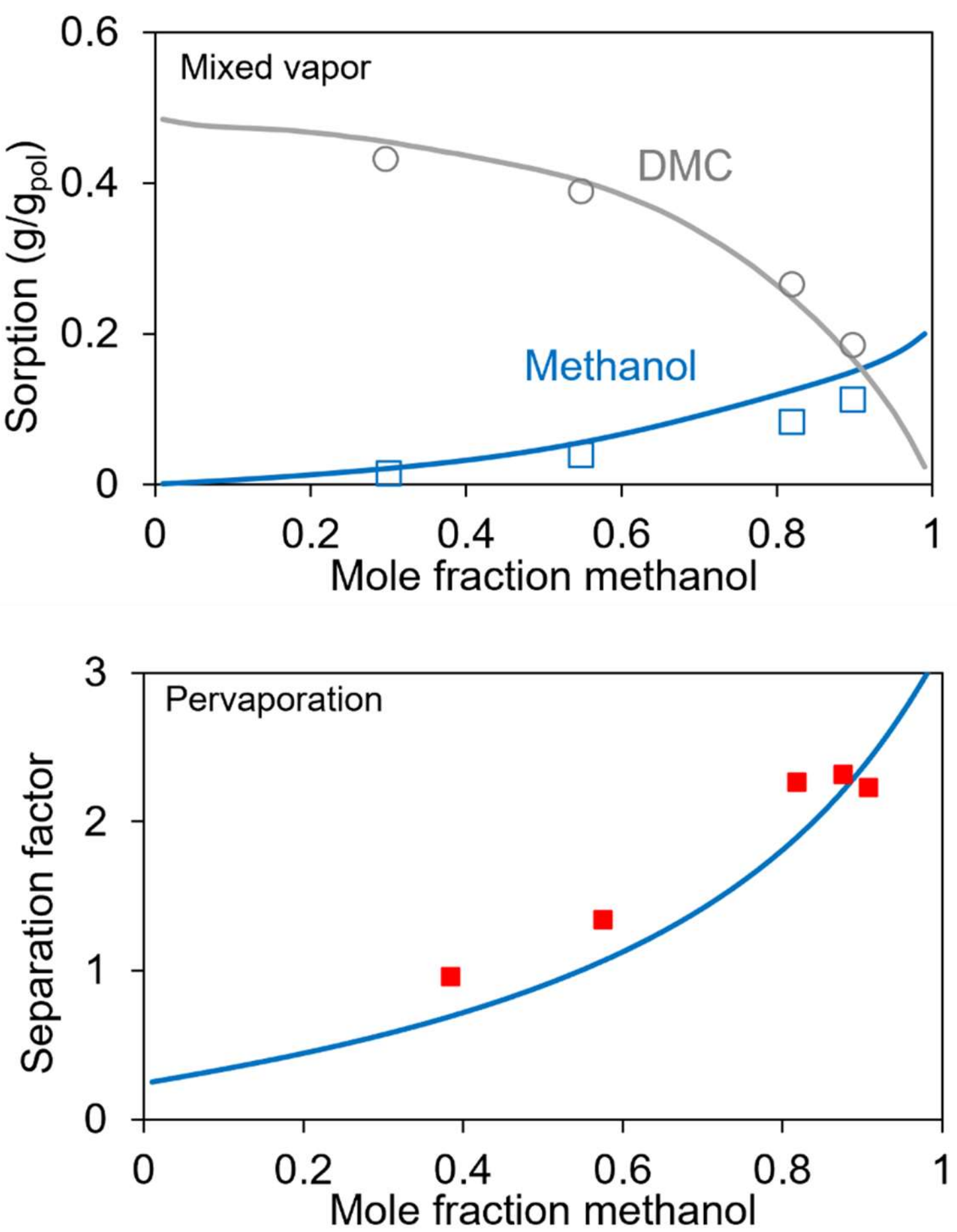

**Figure 6:** Top: Comparison of DGRPT predictions[33] to data[36] for the mixed vapor sorption of DMC / methanol at 40 °C and total pressure fixed to 48% of the dew point pressure. Bottom: Comparison of Maxwell-Stefan diffusion + DGRPT model prediction[33] to data[36] for the membrane pervaporation of methanol / DMC liquid mixtures at 40 °C.

**SBAD-1**

SBAD-1[38] is a spirocyclic polymer with N-aryl bonds; glassy microporous SBAD-1 membranes were demonstrated to fractionate crude oil on the basis of molecular weight and class. Marshall et al.[8] fit pure SBAD-1 PC-SAFT parameters simultaneously to octane and toluene vapor sorption data. With the pure polymer parameters fixed, parameterization of additional hydrocarbon species was achieved by adjusting that species binary interaction parameter with the polymer $k_{i,p}$ to pure vapor sorption data. In order to apply the developed model to petroleum, Marshall et al.[39] proposed a simple correlation for $k_{i,p}$ which depends on the structure of the hydrocarbon.

Figure 7 shows a comparison between model and data for pure vapor sorption of 9 different hydrocarbons. DGRPT gives a good representation of the sorption in each case. All hydrocarbons with the exception of 1MN exhibit Langmuirian type sorption behavior. The exception to this rule is 1MN, whose isotherm exhibits a convex character.

Marshall et al.[40] incorporated this DGRPT into a membrane diffusion model to predict OSRO separations of complex hydrocarbon mixtures. Taking the pure component limit of a mole fraction based Maxwell-Stefan[37] diffusion model for the flux $N_i$ from a high-pressure liquid retentate to a low-pressure liquid permeate,

$$N_i = -\frac{\rho_i}{k_B T} \frac{Đ_{i,p}}{x_p} \frac{d\mu_i}{dz} \tag{15}$$

$\rho_i$ is the guest species density in the polymer membrane, $Đ_{ip}$ is the Maxwell-Stefan diffusivity of component $i$ in the membrane and $x_p$ is the polymer mole fraction in the membrane.

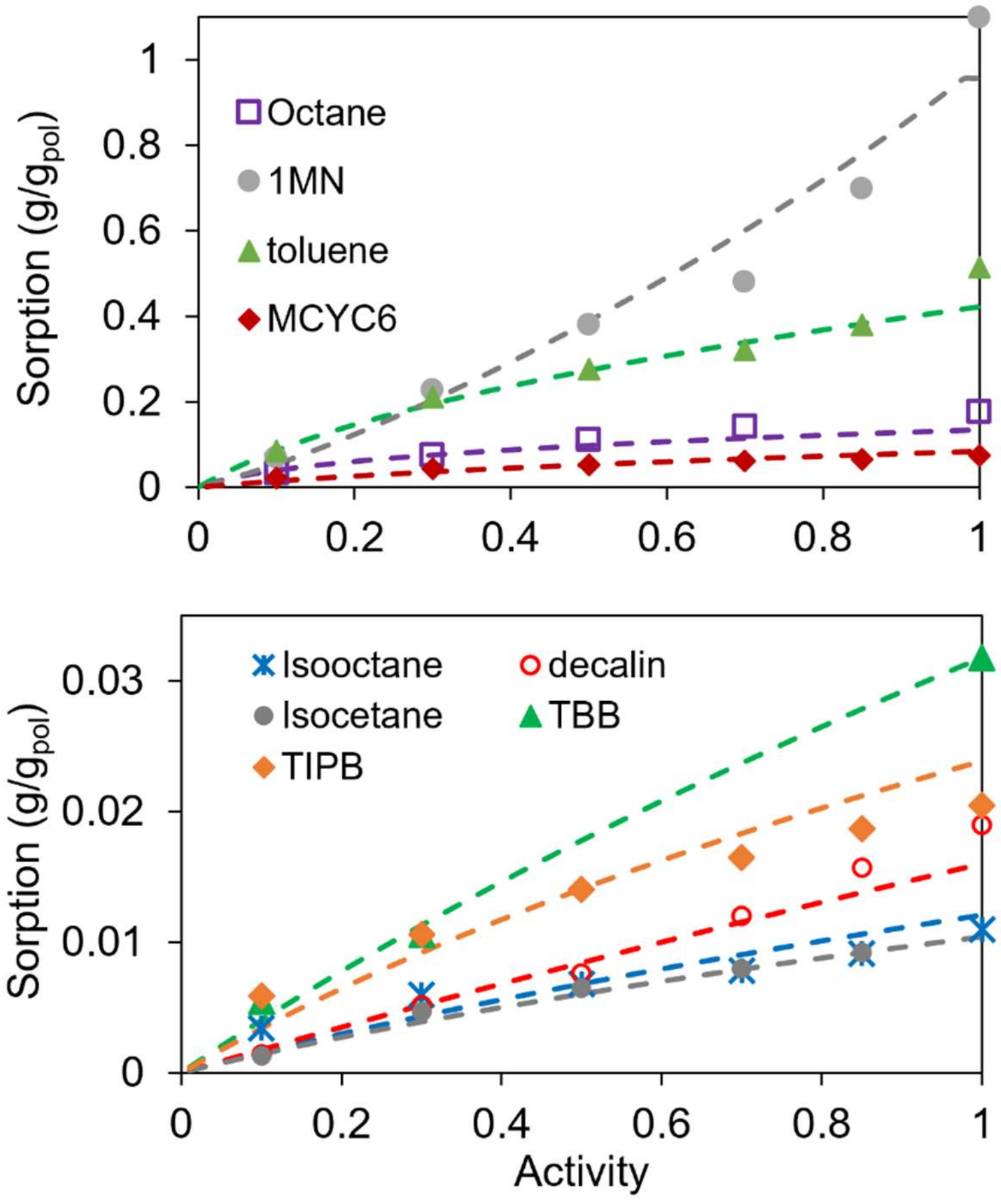

**Figure 7:** Comparison of DGRPT predictions[8] (curves) to data[41] (symbols) for sorption of hydrocarbons on SBAD-1 at $T$ = 25 ˚C. MCYC6 = methylcyclohexane, TBB = tert-butyl benzene, TIPB = 1,3,5-triisopropylbenzene, 1MN = 1-methylnaphthalene

The membrane model is applied by employing DGRPT to calculate the fluid species solubilities at the permeate and retentate boundaries, and then regressing $Đ_{ip}$ in Eq. (15) to reproduce permeation of pure liquids. The results of this fitting can be seen in Fig. 8. In all cases the permeate is at 1 atm of absolute pressure. As can be seen, the flux verse pressure curves of toluene, octane, MCYC6, 1MN exhibit a negative curvature. All remaining species exhibit a linear pressure dependence or even positive curvature.

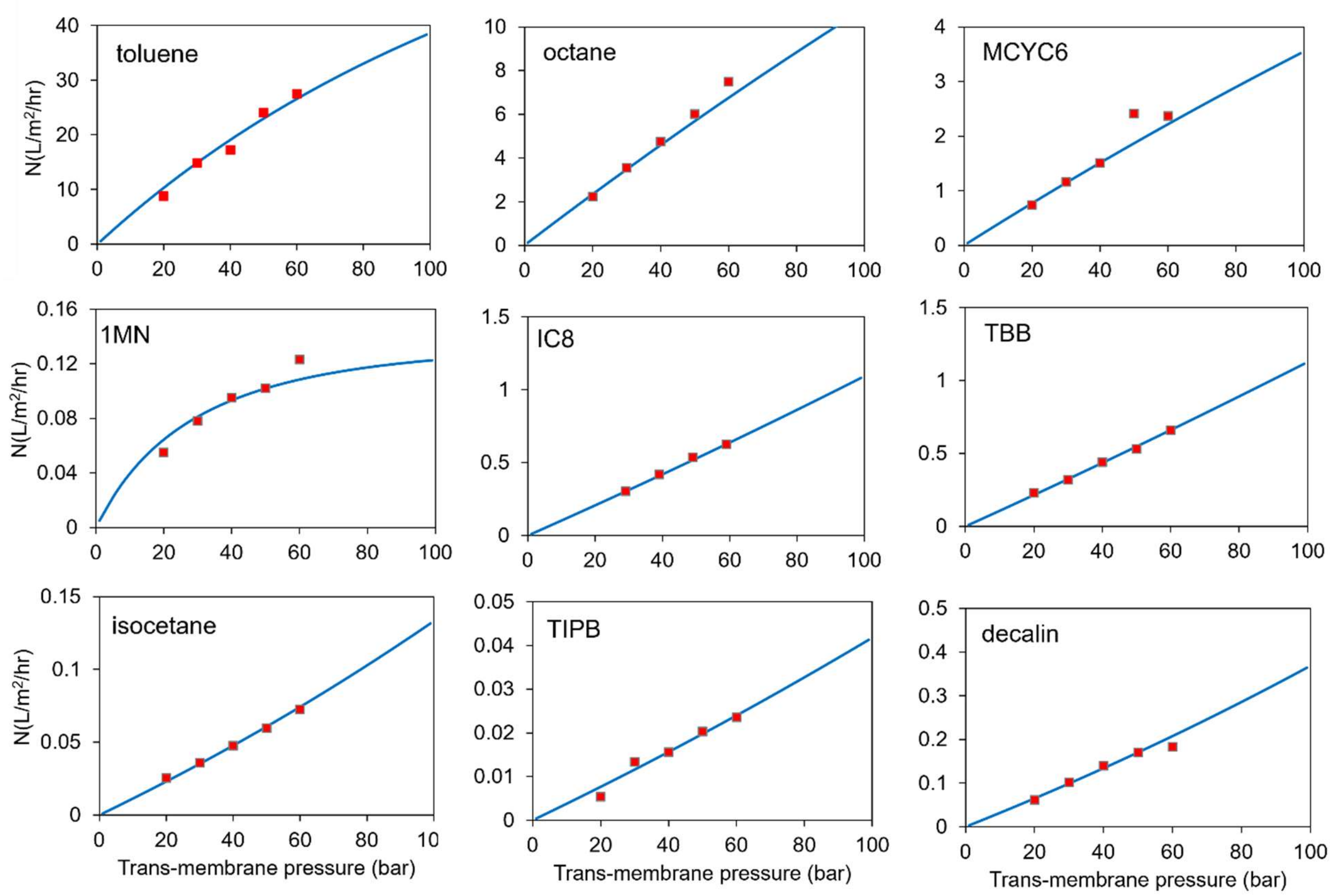

**Figure 8:** Comparison of model[40] and data (points)[8] for the flux of pure hydrocarbon liquids through SBAD-1 at 22 °C.

The variations in pressure dependence can be explained by DGRPT predictions of solubilities as illustrated in Fig. 9. Fluid species which exhibit negative curvature in the flux vs. pressure relation have a higher solubility in SBAD-1. At low pressure this results in highly swollen membranes. As the pressure increases, the membrane is compressed, resulting in a decrease in the liquid solubility. This decrease in solubility due to compression manifest in Fig. 8 as negative curvature in the flux vs. pressure curves. On the other hand, fluid species which exhibit positive curvature in the flux vs. pressure curves have low solubility in SBAD-1. Since at low pressure the membrane is non-swollen, increasing pressure has the effect of increasing solubility in SBAD-1. This manifest as positive curvature in the flux vs. pressure curves.

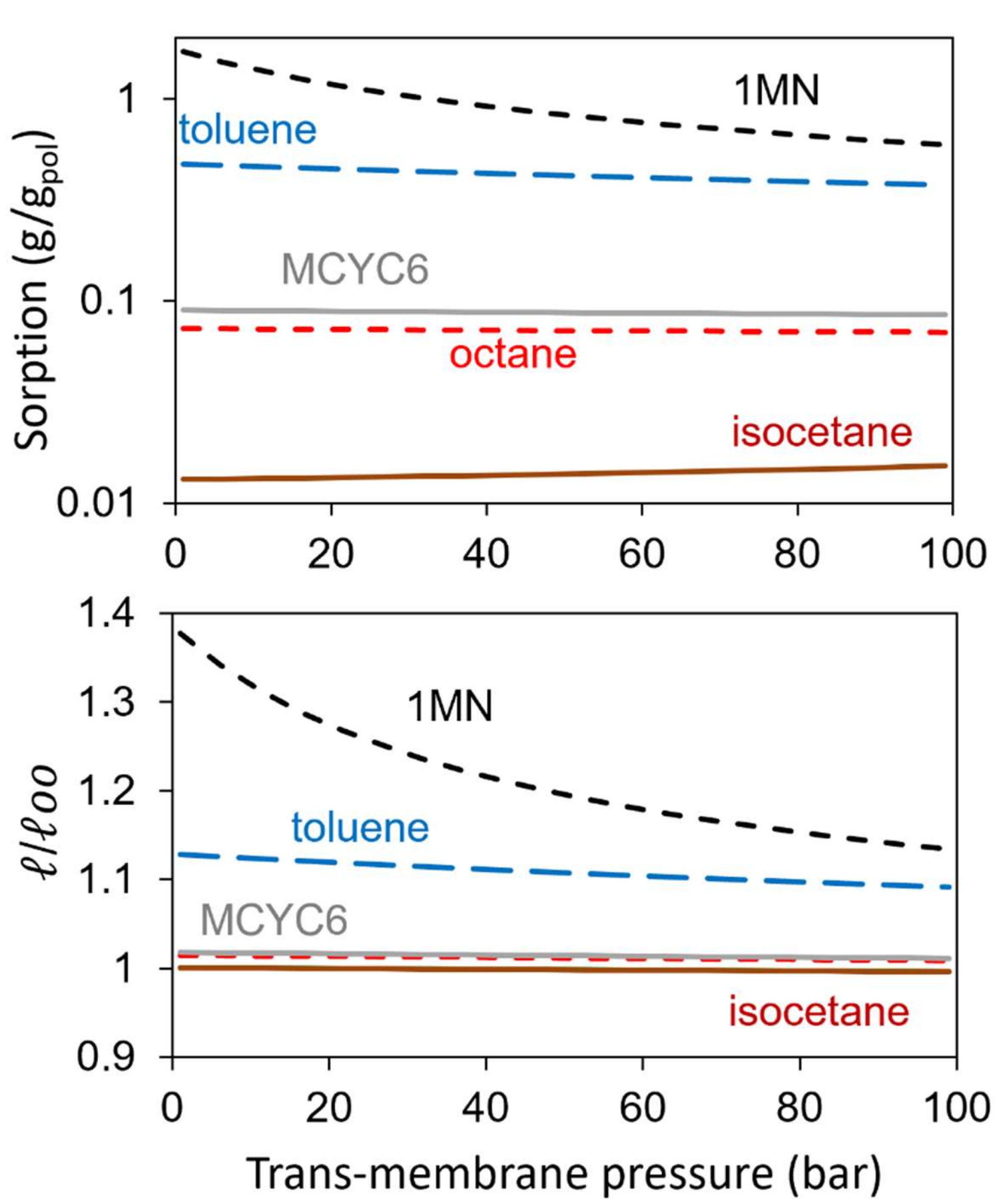

**Figure 9:** DGRPT predictions[40] of sorption and membrane thickness for several pure liquids permeating through an SBAD-1 membrane. $\ell_{oo}$ is the dry membrane thickness at atmospheric pressure.

It was demonstrated by Marshall et al.[40] that a diffusivity averaging approach could be used to predict the OSRO separation of hydrocarbon mixtures in several glassy polymer membranes.

$$N_i = -\frac{\rho_i}{k_B T}\frac{Đ_{avg}}{x_p}\frac{d\mu_i}{dz} \tag{16}$$

where $Đ_{avg}$ is the compositionally averaged diffusivity within the membrane. Employing Eq. (16), it is easily demonstrated that the composition of the permeate liquid can be predicted without any knowledge of diffusivities. Hence, with nothing but a fluid phase equation of state to predict driving force, and a solubility model, the membrane performance in terms of separation factor can be predicted, in the total absence of permeation data. Of course, this is only true if the mixture diffusion is fully mutualized as described in Eq. (16).

Marshall et al.[39] applied this approach to predict the OSRO separation of a light Permian crude with an SBAD-1 membrane. A detailed model of composition (MOC) based on structure oriented lumping (SOL)[42, 43] was available for this crude oil. An MOC provides a detailed composition of petroleum based on 1000's of individual hydrocarbon species. MOC's are constructed through extensive characterization using advanced methods such as 2D-GC. This MOC provided the compositional basis of the petroleum used in the DGRPT predictions, with individual hydrocarbon PC-SAFT parameters estimated using the PC-SAFT petroleum parameterization of Marshall et al.[44]

The top panel of Fig. 10 compares model prediction to data for the permeate simulated distillation curve obtained in the OSRO separation. The model is in good agreement with the data. The data at low molecular weights is not reliable due to loss of light ends in the measurement process. The center panel gives model predictions of the selectivity $R_i = x_i^{perm}/x_i^{ret}$ of molecules in several homologous series as a function of molecular weight. For all homologous series, at low molecular weights the model predicts increasing $R_i$ with increasing molecular weight, $R_i$ then goes through a maximum and then begins to decrease with increasing molecular weight. This complex behavior results from a trend of increasing solubility with increasing molecular weight for small molecules, and decreasing solubility with increasing molecular weight for large molecules. This represents a pure theoretical prediction not previously described in the literature. There is supporting experimental evidence.

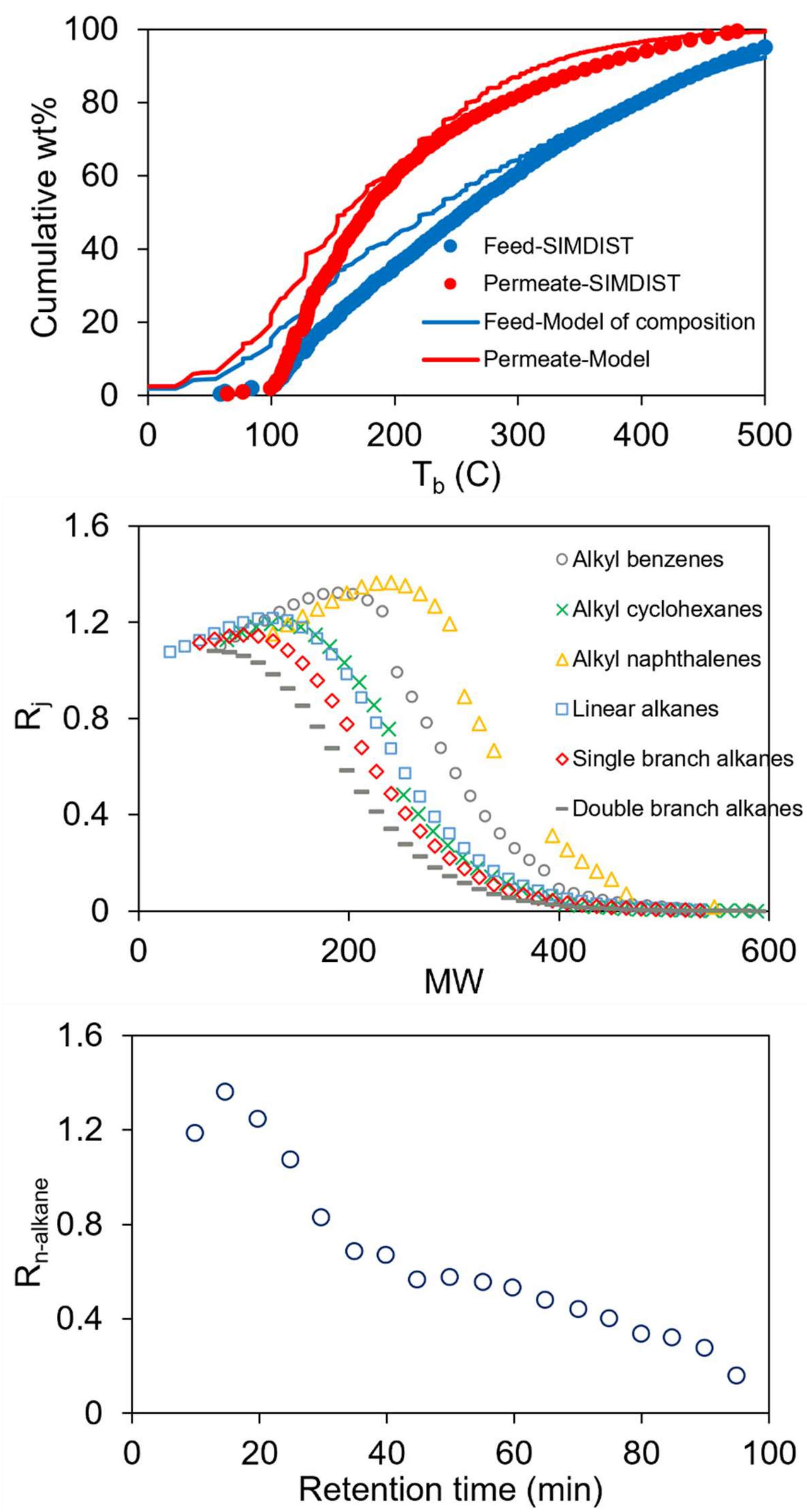

**Figure 10:** Model predictions[39] and experimental results[4] for the OSRO separation of light Permian crude oil at 130 °C and 55 bar using a SBAD-1 membrane. Top: Comparison of model and date for the simulated distillation of the permeate. Middle: Model predicts of selectivity for several homologous series. Bottom: 2D-GC measurement[4] of alkane selectivity

The bottom panel of Fig. 10 gives $R_i$ measurements using 2D-GC for the n-alkane series as a function of GC retention time. Retention time is strongly correlated with molecular weight. As can be seen, $R_i$ initially increases with molecular weight, goes through a maximum and then begins decreasing. This data provides experimental support for the predicted molecular weight dependence of $R_j$ in the center panel of Fig. 10.

DGRPT predictions of the OSRO separation of petroleum with SBAD-1 membranes was further applied to membrane cascades upstream of a crude distillation unit.[45] It was shown that certain membrane cascade configurations could decrease energy usage of the separation by up to 44%.

**Matrimid®5218**

DGRPT was applied[46] to predict the OSRO separations of alcohol / hydrocarbon mixtures using Matrimid membranes. The pure component PC-SAFT parameters $m$, $\sigma$ and $\varepsilon$ where fit to reproduce the pure vapor sorption of acetone and methyl-acetate.[39] Additional guest species were parameterized by fitting a binary interaction parameters $k_{i,p}$ between the guest species and polymer. Figure 11 compares model predictions of hydrocarbon vapor sorption, polymer swelling and alcohol vapor sorption. As can be seen, the model is highly accurate in the description of hydrocarbon sorption over a range of molecular classes. DGRPT gives a good representation of ethanol and n-pentanol sorption, although not as accurately as hydrocarbons. It should be noted that the hydrogen bond volume and energy of Matrimid were not adjusted to sorption data. Instead, these parameters were fixed using the carbonyl oxygen group association parameters of Marshall and Bokis[47].

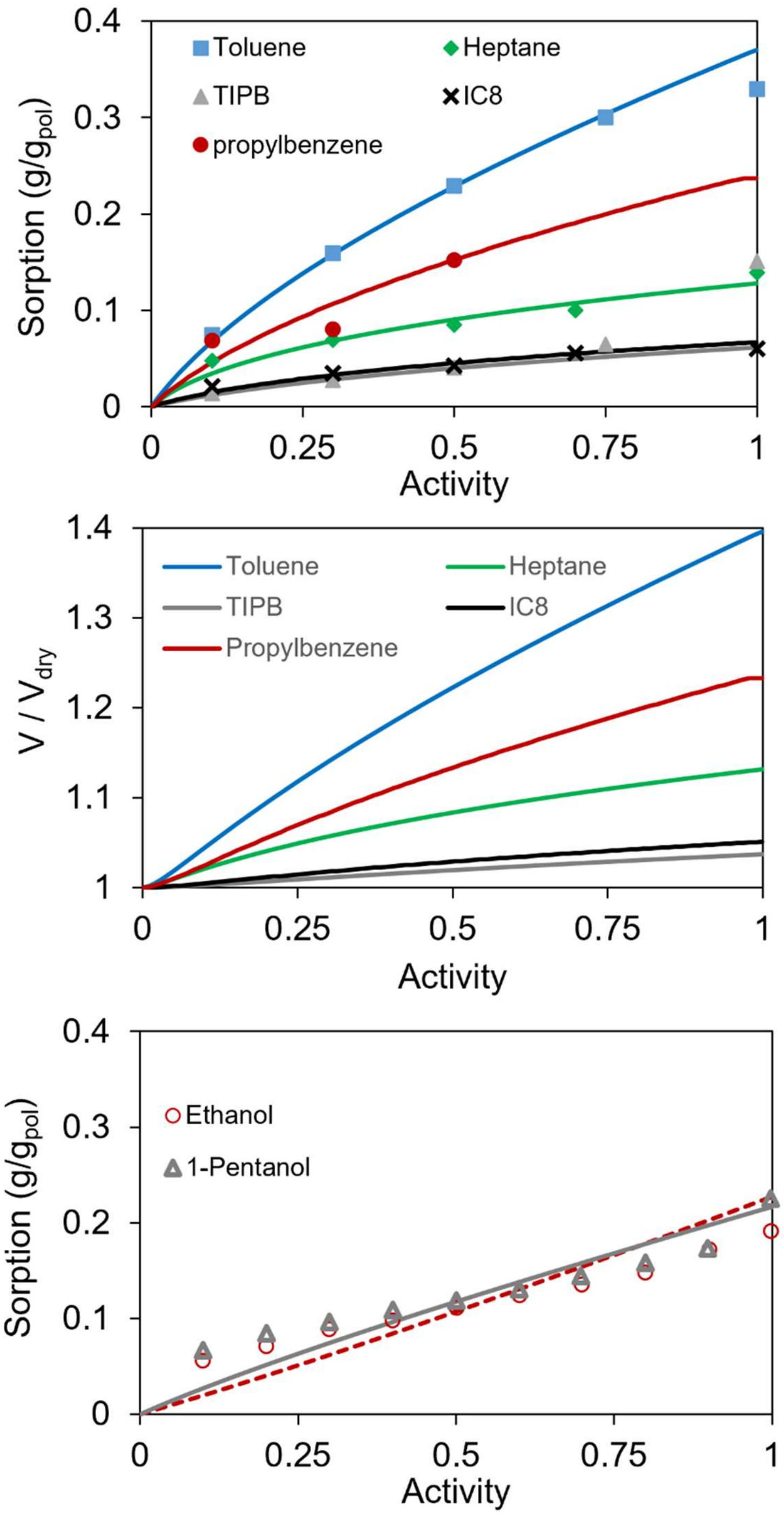

**Figure 11:** Top: Comparison of DGRPT predictions[46] to hydrocarbon sorption data[46] at 25 °C. Data for triisopropylbenzene (TIPB), isooctane (IC8) and propylbenzene are unpublished.[48] Binary interaction parameters for these three species were fit in the current work: TIPB(0.16), IC8(0.16), propylbenzene(0.082). Center: DGRPT predictions of polymer swelling. Bottom: Comparison of DGRPT predictions to Ethanol and pentanol sorption data[49] at a temperature of 30 °C.

Marshall et al.[46] modelled the OSRO separation of hydrocarbon / alcohol mixtures using a simple Fickian diffusion theory with membrane boundary solubilities calculated with DGRPT.

$$N_i = D_i \frac{\rho_i(0) - \rho_i(\ell)}{\ell} \tag{17}$$

Figure 12 illustrates the model predictions for the OSRO separation of n-pentanol / toluene and ethanol / heptane liquid mixtures.

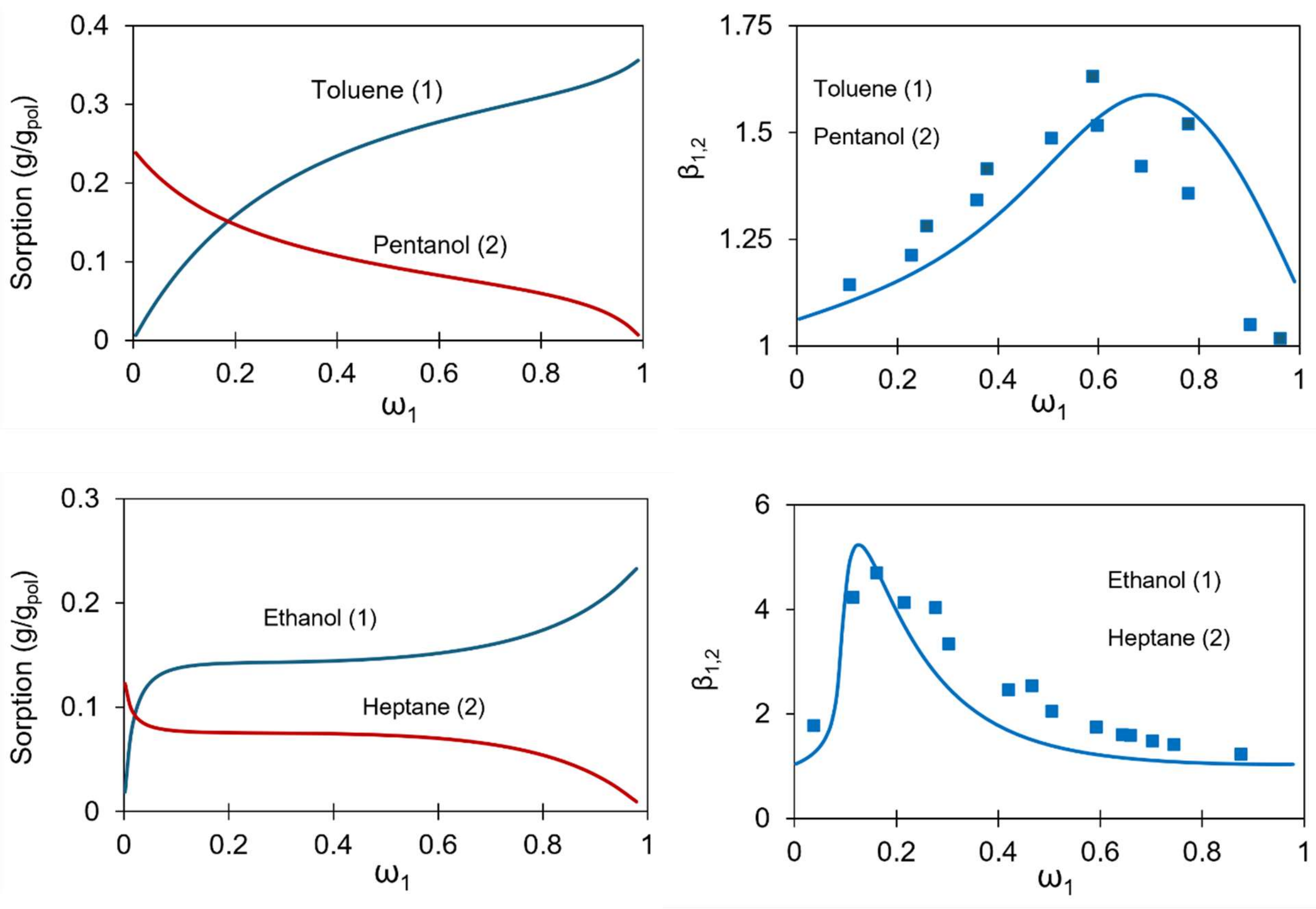

**Figure 12:** Mixture sorption and separation factors for the OSRO separation of toluene / pentanol (top) and ethanol / heptane (bottom) liquid mixtures at 25 °C. In each case, the retentate pressure is 40 bar and the permeate pressure is 1 bar.[46] The independent variable in each graph is the mass fraction $\omega_1$ of species 1 in the retentate.

The left panels of Fig. 12 give the DGRPT sorption predictions from binary liquid mixtures at a pressure of 40 bar. The sorption is highly non-ideal. The right panel gives the separation factors for the OSRO separation of these mixtures. The separation factors are highly non-linear, exhibiting extrema as a function of retentate composition. Since the diffusion is Fickian, this complex compositional dependence is solely the result of non-idealities related to liquid sorption at the membrane boundaries. DGRPT accurately predicted these separation factors, using only pure vapor sorption data to parameterize the model. This highlights the power of the DGRPT approach for the prediction of OSRO separations with deep glassy membranes.

## V: Extensions of DGRPT

Section IV demonstrated the wide applicability of the first order DGRPT expansion for high free volume deep glassy polymers. When considering non-hydrogen bonding fluid species / polymers, the first order DGRPT expansion best represents sorption isotherms with a Langmuirian character. However, it is common for dense glassy polymers to exhibit a dual mode[50] sorption behavior. For a pure gas the dual mode isotherm is given by

$$q = q_{Lang} + q_{lin} = \frac{AP}{1 + BP} + CP \tag{18}$$

Where $q$ is the sorption, $q_{Lang}$ is the Langmuir adsorption, $q_{lin}$ is the linear sorption and $P$ is the pressure. When $C = 0$, Langmuir sorption is recovered. Langmuir isotherms saturate at high pressure obeying the limit

$$q_{Lang}(P \longrightarrow \infty) = \frac{A}{B} \tag{19}$$

Hence, the major assumptions 3) and 4) of DGRPT hold for Langmuir type sorption behavior. The dry glass reference state is imprinted on the polymer. Linear isotherms diverge in the high pressure limit,

$$q_{lin}(P \longrightarrow \infty) = \infty \tag{20}$$

Linear type sorption is unbounded. They dry glass reference state is destroyed. The polymer is essentially soluble in the gas. For this type of system, a first order expansion as in Eq. (7) is not sufficient.

**Second Order DGRPT**

To increase the accuracy for systems which exhibit linear type sorption, Ismael et al[51] extended DGRPT to include a second order term. The second order DGRPT expansion is given by

$$\mu_p\left(\rho_p,\{\rho_j\},T\right) \approx \mu_p\left(\rho_p^o,T\right) + \sum_{i=1}^{n}\left(\left.\frac{\partial \mu_p\left(\rho_p^o,\{\rho_j\},T\right)}{\partial \rho_i}\right|_{\{\rho_j\}=0} \rho_i + \frac{\gamma_i}{2}\left.\frac{\partial^2 \mu_p\left(\rho_p^o,\{\rho_j\},T\right)}{\partial \rho_i^2}\right|_{\{\rho_j\}=0} \rho_i^2\right) \quad (21)$$

$\gamma_i$ is a combined variable given by

$$\gamma_i = 1 + \frac{\left.\frac{\partial \mu_p\left(\rho_p^o,\{\rho_j\},T\right)}{\partial \rho_p}\right|_{\{\rho_j\}=0}}{\left.\frac{\partial^2 \mu_p\left(\rho_p^o,\{\rho_j\},T\right)}{\partial \rho_i^2}\right|_{\{\rho_j\}=0}} \left.\frac{\partial^2 \rho_p}{\partial \rho_i^2}\right|_{\{\rho_j\}=0} \quad (22)$$

The unknown term in Eq. (22) is the second derivative of the polymer density at infinite dilution. The first order expansion is recovered when the following holds true,

$$\left.\frac{\partial^2 \rho_p}{\partial \rho_i^2}\right|_{\{\rho_j\}=0} = -\frac{\left.\frac{\partial^2 \mu_p\left(\rho_p^o,\{\rho_j\},T\right)}{\partial \rho_i^2}\right|_{\{\rho_j\}=0}}{\left.\frac{\partial \mu_p\left(\rho_p^o,\{\rho_j\},T\right)}{\partial \rho_p}\right|_{\{\rho_j\}=0}} \quad (23)$$

There is no clear way to estimate the second derivative of the polymer density *a priori*. Future work should focus on developing methodology to predict this second derivative from polymer and fluid species properties. Ismaeel et al. treated the combined variable $\gamma_i$ is treated as an adjustable parameter. When applied in this way, second order DGRPT is a two parameter model, similar to the original NETGP with closure given by Eq. (3). That said, DGRPT is built on a stronger theoretical foundation. Equation (3) refers only to fluid phase properties, while Eq. (21) is described by polymer phase properties. This distinction may be important in multi-component mixtures where there are strong competitive or cooperative effects between sorbed fluid molecules in the polymer phase.

For instance, consider the following thought experiment about sorption from a binary mixture of component 1 and component 2. Assume Eq. (3) had been parameterized to pure vapor sorption experiments of component 1 in the polymer and then component 2 in the polymer to yield $k_{s,1}$ and $k_{s,2}$. Now imagine vapor sorption in an equimolar binary mixture of components 1 and 2 such that, for whatever reason, component 2 outcompetes component 1 and completely excludes component 1 from sorbing into the polymer. Equation (3) would then predict the polymer swelling to be

$$\frac{\rho_p}{\rho_p^o} = \frac{1}{1 + \frac{P}{2}\left(k_{s,1} + k_{s,2}\right)} \tag{24}$$

Equation (24) is clearly incorrect as the polymer swelling depends on $k_{s,1}$ even though component 1 is not present in the polymer phase. Now evaluating second order DGRPT for the same contrived system

$$\mu_p\left(\rho_p, \{\rho_j\}, T\right) \approx \mu_p\left(\rho_p^o, T\right) + \left.\frac{\partial \mu_p\left(\rho_p^o, \{\rho_j\}, T\right)}{\partial \rho_2}\right|_{\{\rho_j\}=0} \rho_2 + \frac{\gamma_2}{2} \left.\frac{\partial^2 \mu_p\left(\rho_p^o, \{\rho_j\}, T\right)}{\partial \rho_2^2}\right|_{\{\rho_j\}=0} \rho_2^2 \tag{25}$$

As can be seen, the chemical potential of the polymer only depends on the derivatives of component 2. This is because DGRPT references only polymer phase variables, while Eq. (3) references fluid phase variables. Hence, in principle, DGRPT could predict the effect of competitive sorption on polymer swelling for this system, while Eq. (3) cannot.

Second order DGRPT in now applied to toluene / PIM-1 sorption at 25 ◦C shown n Fig. 5. PIM-1 parameters are taken from Marshall and Johsnon[33], and the binary interaction parameter $k_{i,p}$ = 0.0354 and swelling parameter $\gamma_i$ = -0.03 were adjusted to the data. The comparison is given in Fig. 12. As can be seen, the second order DGRPT gives a better description of the 25 ◦C data. However, as can be observed in Fig. 5, the first order DGRPT is quantitatively accurate at 55 ◦C. The difference between the two temperatures, is that the 25 ◦C exhibits a dual mode character, while 55 ◦C is purely Langmuirian.

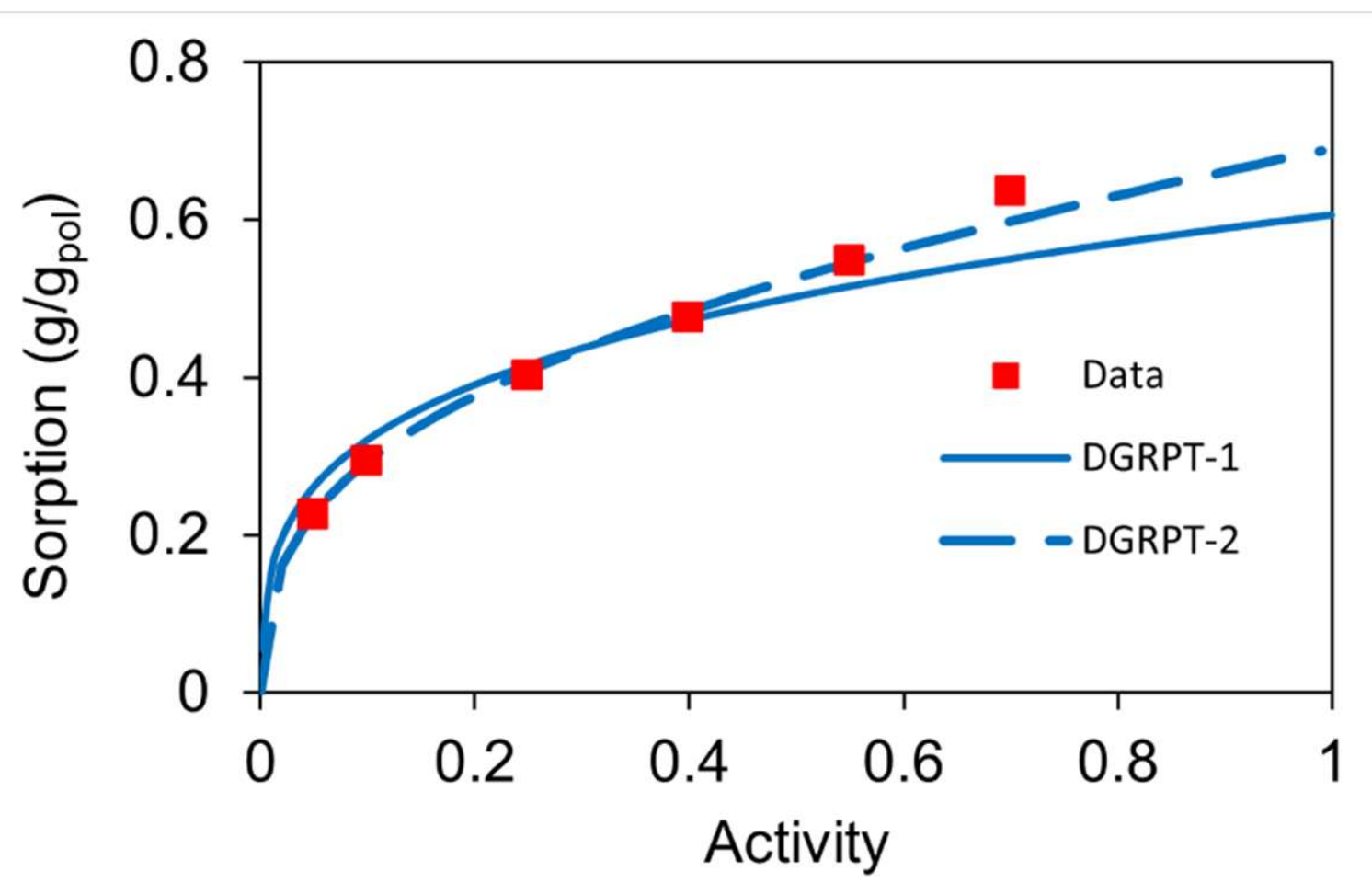

**Figure 12:** Comparison of first order DGRPT[33] (solid curve), second order DGRPT (dashed curve) and experimental data[35] for the sorption of toluene in PIM-1 at 25 ◦C.

**A multi-density expansion:**

In an attempt to extend first order DGRPT to linear sorption, Marshall proposed the following modified expansion of the polymer chemical potential[52]

$$\mu_p(\rho_p,\rho_i,T) \approx \mu_p(\rho_p^o,T) + \left.\frac{\partial \mu_p(\rho_p^o,\rho_i,T)}{\partial \rho_i}\right|_{\rho_i=0} \rho_i + \left.\frac{\partial \mu_p(\rho_p,\rho_i,T)}{\partial \rho_p}\right|_{\rho_i=0} (\rho_p - \rho_p^o) \qquad (26)$$

Equation (26) expands around the dry glass state with respect to both the guest species density and the polymer density. This version of DGRPT was shown to accurately represent $CO_2$ sorption in PMMA[52], where the standard first order DGRPT failed, as well as reproduce equilibrium sorption in semi-crystalline polymers.

It was later noted by Ismaeel et al.[51], that this version of DGRPT results in two plausible solutions to the sorbed density. The genesis of multiple solutions is the fact that the polymer density appears on the right hand side of Eq. (26). Under further study, which we conduct here, we note that one of the solutions has a glassy character, while the other solution has an equilibrium character. Figure 13 compares model predictions to data for CO2 sorption in PMMA. The PMMA PC-SAFT parameters of the original study[52] are used in the calculations. The equilibrium type root gives a convex isotherm which accurately describes the data, while the glassy type root gives a concave isotherm which is not in agreement with the data. An interesting future research path could be to attempt to formulate an expansion where a glassy root and equilibrium root cross at the sorption induced glass transition point.

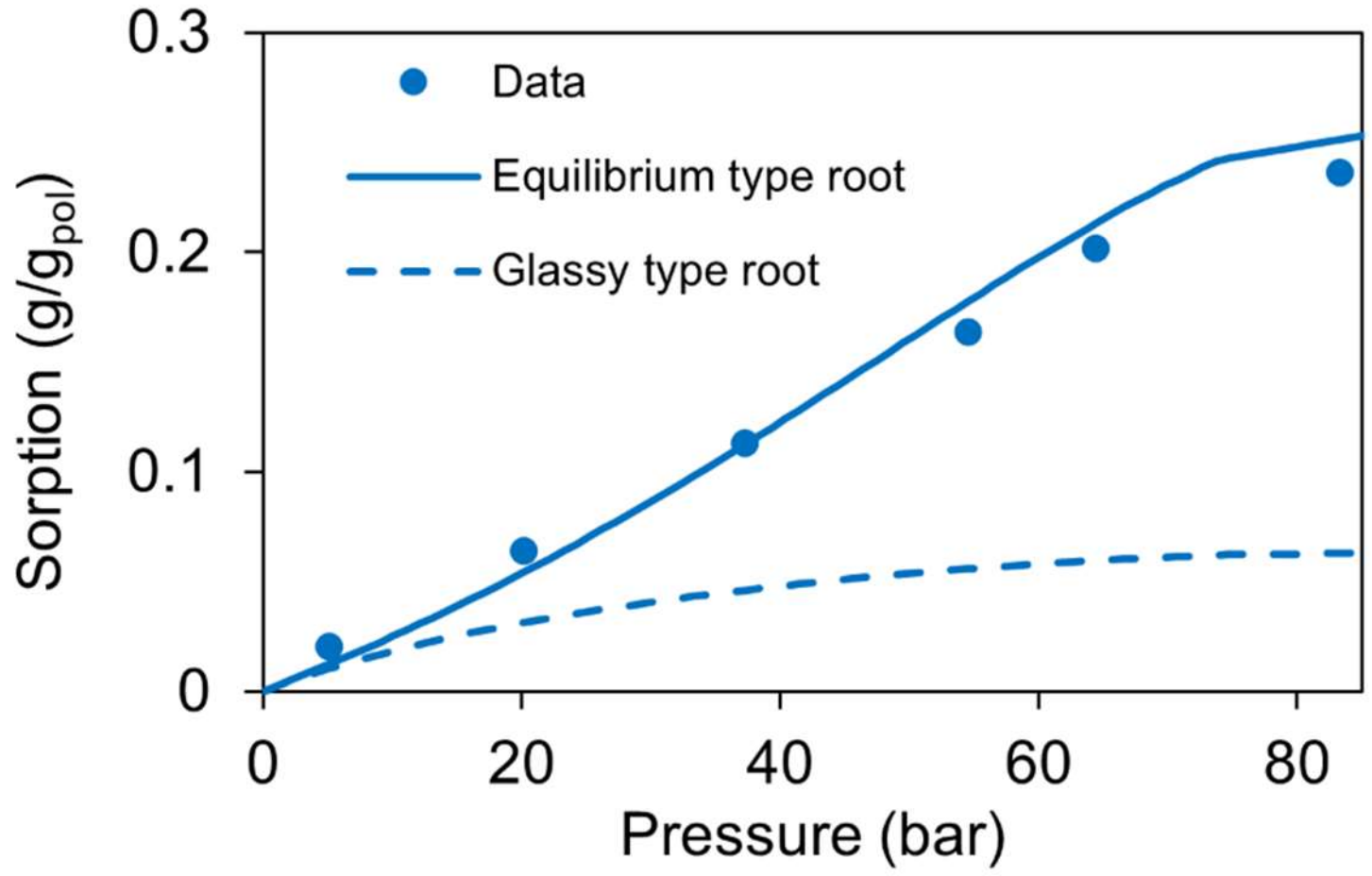

**Figure 13:** Comparison of modified DGRPT[52] using closure Eq. (24) to data[10] for the sorption of $CO_2$ in PMMA at 33 °C. Polymer PC-SAFT parameters and binary interaction parameter ($k_{i,p}$ = 0.0284) same as ref[[52]]

Unlike the closure Eq. (21), the closure Eq. (26), does not introduce any additional parameters. Marshall proposed the following generalization of Eq. (26),

$$\mu_p\left(\rho_p,\rho_i,T\right) \approx \mu_p\left(\rho_p^o,T\right) + \left.\frac{\partial\mu_p\left(\rho_p^o,\rho_i,T\right)}{\partial\rho_i}\right|_{\rho_i=0}\rho_i + \lambda\left.\frac{\partial\mu_p\left(\rho_p,\rho_i,T\right)}{\partial\rho_p}\right|_{\rho_i=0}\left(\rho_p-\rho_p^o\right) \qquad (27)$$

Where $\lambda$ is an adjustable parameter between 0 – 1 which modulates the term which is linear in polymer density.

To eliminate the requirement of this additional fitting parameter, Baldanza et al.[53] proposed the following relationship for λ

$$\lambda = \frac{v_{cp}^{mol}}{v^{mol}} \qquad (28)$$

In Eq. (28) $v^{mol}$ is the molar volume of the polymer phase including guest species, and $v_{cp}^{mol}$ is the close packed molar volume given by the relationship

$$v_{cp}^{mol} = \frac{1}{\sqrt{2}}\left(x_p m_p d_p^3 + \sum_{i=1}^{n} x_i m_i d_i^3\right) \quad (29)$$

Where $x_i$ is the mole fraction in the polymer phase and $d_i$ is the temperature dependent diameter. The combined theory Eq. (1), (27) – (29) again has only a single adjustable parameter $k_{i,p}$. The updated theory was shown to give improved predictions for methanol sorption in PEI and 6FDA-ODA.

## VI: Summary

DGRPT represents a general formalism to describe multi-component sorption in glassy polymers. When applied at first order, the model does not introduce any additional model parameters over the equation of state parameterization. Once the pure polymer PC-SAFT parameters are known, the sorption of each fluid species is parameterized by the binary interaction parameter with the polymer $k_{i,p}$. For high free volume deep glassy polymers which exhibit Langmuir like sorption isotherms, the first order theory is highly accurate. The theory is applied to OSRO separations by first parameterizing the model to pure vapor sorption and then implementing the DGRPT model in the context of a diffusion theory. DGRPT allowed for the accurate prediction of petroleum separations with SBAD-1, pervaporation of DMC and methanol liquid mixtures in PIM-1, and the OSRO separation of alcohol and hydrocarbon mixtures in Matrimid.

The theory loses accuracy in systems which exhibit sorption which is linearly dependent on gas partial pressure. This is due to the fact that linear sorption isotherms do not saturate, they allow for the complete dissolution of the polymer at high pressure. Clearly, a first order perturbation theory to the dry glass reference state cannot represent this unbounded sorption. Second order DGRPT allows for the more accurate representation of these linear regions, however at second order an additional adjustable parameter $\gamma_i$ is introduced to the theory. Hence the second order form of the theory has the same number of adjustable parameters as the original NETGP approach. That said, second order DGRPT is theoretically capable of predicting the effect of competitive sorption on polymer swelling, while Eq. (3) is not.

Lastly the two density expansion first order expansion of the polymer chemical potential around the dry glass reference state was discussed. It was shown that this expansion gives two roots, a glass like root and an equilibrium type root. An interesting line of future research will be to study the applicability of this approach (suitably modified) to predict sorption induced glass transitions.